\documentstyle[12pt]{article}
\input epsf.tex

\begin{document}

\baselineskip=15.5pt
\pagestyle{plain}
\setcounter{page}{1}

\renewcommand{\thefootnote}{\fnsymbol{footnote}}

\begin{titlepage}

\begin{flushright}
PUPT-1888\\
hep-th/9908165
\end{flushright}
\vfil

\begin{center}
{\huge Absorption by Threebranes and the AdS/CFT Correspondence\footnote{
Talk at {\sl Strings '99}, Potsdam, Germany.}
}
\end{center}

\vfil

\begin{center}
{\large Igor R.\ Klebanov}\\
\vspace{3mm}
Joseph Henry Laboratories\\
Princeton University\\
Princeton, New Jersey 08544, USA\\
\vspace{3mm}
\end{center}

\vfil

\begin{center}
{\large Abstract}
\end{center}

\noindent
In the first part of this talk I discuss two somewhat different
supergravity approaches to calculating correlation functions in 
strongly coupled Yang-Mills
theory. The older approach relates 
two-point functions to cross-sections for absorption of certain
incident quanta by threebranes. In this approach the normalization of
operators corresponding to the incident particles is fixed unambiguously by
the D3-brane DBI action.
 By calculating absorption cross-sections
of all partial waves of the dilaton we find corresponding
two-point functions at strong `t Hooft coupling and show that
they are identical to the weak coupling results. The newer approach 
to correlation functions relates them to boundary conditions in AdS space.
Using this method
we show that for a certain range of negative mass-squared there are
two possible operator dimensions corresponding to a given scalar field
in AdS, and indicate how to calculate correlation functions for either
of these choices.
In the second part of the talk I discuss an example of AdS/CFT
duality which arises in the context of type 0 string theory.
The CFT on $N$ coincident electric and magnetic D3-branes is argued
to be stable for sufficiently weak `t Hooft coupling. It is suggested that
its transition to instability at a critical coupling is related to
singularity of planar diagrams.

\vfil
\begin{flushleft}
August 1999
\end{flushleft}
\end{titlepage}
\newpage

\renewcommand{\thefootnote}{\arabic{footnote}}
\setcounter{footnote}{0}


\newcommand{\grad}{\nabla}
\newcommand{\tr}{\mathop{\rm tr}}
\newcommand{\half}{{1\over 2}}
\newcommand{\third}{{1\over 3}}
\newcommand{\be}{\begin{equation}}
\newcommand{\ee}{\end{equation}}
\newcommand{\bea}{\begin{eqnarray}}
\newcommand{\eea}{\end{eqnarray}}

\newcommand{\dint}[2]{\int\limits_{#1}^{#2}}
\newcommand{\D}{\displaystyle}
\newcommand{\PDT}[1]{\frac{\partial #1}{\partial t}}
\newcommand{\PD}{\partial}
\newcommand{\tw}{\tilde{w}}
\newcommand{\tg}{\tilde{g}}
\newcommand{\newcaption}[1]{\centerline{\parbox{6in}{\caption{#1}}}}
\def\href#1#2{#2}  

\def \ci {\cite}
\def \foot {\footnote}
\def \bi{\bibitem}
\newcommand{\rf}[1]{(\ref{#1})}
\def \del{\partial}
\def \m {\mu}
\def \n {\nu} 
\def \g {\gamma}
\def \G {\Gamma}
\def \a {\alpha}
\def \ov {\over}
\def \la {\label}
\def \ep {\epsilon}
\def \d {\delta}
\def \k {\kappa}
\def \p {\phi}
\def \ha {\textstyle{1\ov 2}}
\def \Tr {{\rm Tr}}
\def \b {\beta}

\def\np {  {\em Nucl. Phys.} }
\def \pl { {\em Phys. Lett.} }
\def \mpl { Mod. Phys. Lett. }
\def \prl { Phys. Rev. Lett. }
\def \pr  { {\em Phys. Rev.} }
\def \cqg { Class. Quantum Grav.}
\def \jmp { Journ. Math. Phys. }
\def\ap { Ann. Phys. }
\def \ijmp { Int. J. Mod. Phys. }

%
\def\TL{\hfil$\displaystyle{##}$}
\def\TR{$\displaystyle{{}##}$\hfil}
\def\TC{\hfil$\displaystyle{##}$\hfil}
\def\TT{\hbox{##}}
\def\seqalign#1#2{\vcenter{\openup1\jot
  \halign{\strut #1\cr #2 \cr}}}

\def\comment#1{}
\def\fixit#1{}

\def\tf#1#2{{\textstyle{#1 \over #2}}}
\def\df#1#2{{\displaystyle{#1 \over #2}}}

\def\mop#1{\mathop{\rm #1}\nolimits}

\def\ad{\mop{ad}}
\def\coth{\mop{coth}}
\def\csch{\mop{csch}}
\def\sech{\mop{sech}}
\def\Vol{\mop{Vol}}
\def\vol{\mop{vol}}
\def\diag{\mop{diag}}
\def\tr{\mop{tr}}
\def\Disc{\mop{Disc}}
\def\sgn{\mop{sgn}}

\def\SU{{\rm SU}}
\def\USp{{\rm USp}}            

\def\lsim{\mathrel{\mathstrut\smash{\ooalign{\raise2.5pt\hbox{$<$}\cr\lower2.5pt\hbox{$\sim$}}}}}
\def\gsim{\mathrel{\mathstrut\smash{\ooalign{\raise2.5pt\hbox{$>$}\cr\lower2.5pt\hbox{$\sim$}}}}}

\def\slashed#1{\ooalign{\hfil\hfil/\hfil\cr $#1$}}

\def\sqr#1#2{{\vcenter{\vbox{\hrule height.#2pt
         \hbox{\vrule width.#2pt height#1pt \kern#1pt
            \vrule width.#2pt}
         \hrule height.#2pt}}}}
\def\square{\mathop{\mathchoice\sqr56\sqr56\sqr{3.75}4\sqr34\,}\nolimits}

\def\idget{$\sqr55$\hskip-0.5pt}
\def\endrow{\hskip0.5pt\cr\noalign{\vskip-1.5pt}}
\def\endyoung{\hskip0.5pt\cr}

\def\href#1#2{#2}  


%
\def\lbldef#1#2{\expandafter\gdef\csname #1\endcsname {#2}}
\def\eqn#1#2{\lbldef{#1}{(\ref{#1})}%
\begin{equation} #2 \label{#1} \end{equation}}
\def\eqalign#1{\vcenter{\openup1\jot
    \halign{\strut\span\TL & \span\TR\cr #1 \cr
   }}}
\def\eno#1{(\ref{#1})}

\def\rmax{{r_{\rm max}}}
\def\gone#1{}

\section{From absorption cross-sections to two-point correlators}

The search for exact relations between gauge fields and strings is an
old and important problem \cite{GT,Sasha}.
A simple idea behind some of the recent advances in this direction 
\cite{jthroat,US,EW}
is that a stack of $N$ coincident D3-branes may be described in two
different ways. In the D-brane formalism \cite{Dnotes}
such a stack is described
by ${\cal N}=4$ supersymmetric four-dimensional $U(N)$ gauge
theory \cite{Witten} which at
low energies interacts weakly with the bulk closed string excitations.
On the other hand,
if $N$ is large, then this stack is a heavy object embedded into a theory
of closed strings which contains gravity. Naturally, this macroscopic
object will curve space: it may be described by classical metric
and other background fields including the Ramond-Ramond 
4-form potential \cite{hs}.

Thus, we have two different descriptions of the stack of D3-branes:
one in terms of the $U(N)$ supersymmetric gauge theory on its world volume,
and the other in terms of the 
R-R charged $3$-brane background of the type IIB
closed superstring theory. 
An early indication that these two descriptions are
indeed equivalent in some appropriate limit was the calculation
of absoprtion cross-sections for low-energy waves incident on the
stack from the transverse directions \cite{kleb,gukt}.

To calculate the absorption cross-sections in the D-brane formalism one
needs the low-energy world volume action for coincident D-branes coupled to
the massless bulk fields. Luckily, these couplings may be deduced
from the D-brane Born-Infeld action. For example, the coupling
of D3-branes to the dilaton $\Phi$, the Ramond-Ramond scalar $C$,
and the graviton $h_{\alpha\beta}$ is given by \cite{kleb,gukt}
\be \label{sint}
   S_{\rm int} = {\sqrt \pi\over\kappa}
\int d^4 x \, \bigg[ \tr \left(
 \tf{1}{4}   {\Phi} F_{\alpha\beta}^2  -
  \tf{1}{4}   {C} F_{\alpha\beta} \tilde {F}^{\alpha\beta} \right)
+ \tf{1}{2} h^{\alpha\beta} T_{\alpha\beta} \bigg] \ ,
 \ee
 where $T_{\alpha\beta}$ is
the stress-energy tensor of the ${\cal N}=4$ SYM theory.

Consider, for instance, absorption of a dilaton incident on the 3-brane
at right angles with a low energy
$\omega$. Since the dilaton couples to ${1\over 4}\tr F_{\alpha\beta}^2$
it can be converted into a pair of back-to-back gluons on the world volume.
The leading order calculation of the cross-section
for weak coupling gives \cite{kleb}
\be\label{absorb}
   \sigma = {\kappa^2  \omega^3 N^2\over 32 \pi} \ ,
 \ee
where $\kappa=\sqrt{8\pi G}$ is the 10-dimensional gravitational 
constant (the factor $N^2$ comes from the degeneracy of
the final states which is the number of different gluon species).

This result was compared with the absorption cross-section by
the extremal 3-brane geometry,
\be
\label{geom}
ds^2 = \left (1+{L^4\over r^4}\right )^{-1/2}
\left (- dt^2 +dx_1^2+ dx_2^2+ dx_3^2\right )
+ \left (1+{L^4\over r^4}\right )^{1/2}
\left ( dr^2 + r^2 d\Omega_5^2 \right )\ .
\ee
The singularity at $r=0$ is a coordinate artefact: in fact, the geometry
is completely non-singular \cite{gt}. Indeed, the limiting form of the
metric as $r\rightarrow 0$ is
\be \label{adsmetric}
ds^2 = {L^2 \over z^2} \left( -dt^2 + d\vec{x}^2 + dz^2 \right) +
    L^2 d\Omega_5^2 \ ,
\ee
where $z={L^2\over r}\gg L$, which describes the space
$AdS_5\times S^5$ with equal radii of curvature $L$.
Thus, the 3-brane geometry may be viewed as a semi-infinite throat 
of radius $L$ which for
$r \gg L$ opens up into flat $9+1$ dimensional space.
Waves incident from the $r \gg L$ region partly reflect back and
partly penetrate into the the throat region $r \ll L$.
The relevant $l$-th partial wave radial equation turns out to be \cite{kleb}
\be
\label{partialthree}
\left [{d^2\over d \rho^2} - {l(l+4) + 15/4\over \rho^2}
+1 + {(\omega L)^4\over \rho^4} \right ] \psi_l(\rho) =0\ ,
\ee
where $\rho = \omega r$. For a low energy $\omega \ll 1/L$ we find
a high barrier separating the two asymptotic regions.
The low-energy behavior of the tunneling probability may be calculated
by the so-called matching method, and the resulting $s$-wave absorption
cross-section is \cite{kleb}
$
\sigma^{l=0}_{SUGRA}= {\pi^4\over 8}\omega^3 L^8 $.
Using the relation
between the radius of the throat and the number of D3-branes
 \cite{gkp},
\be\label{throatrel}
L^4 = {\kappa\over 2\pi^{5/2}} N \ ,
\ee
we find that the supergravity
cross-section agrees exactly with the D-brane one.

In order to get a deeper understanding into why these two very different
calculations produce exact agreement, let us first examine their
range of validity.
 Since $\kappa\sim g_{st}
\alpha'^2$, (\ref{throatrel}) gives
$ L^4 \sim N g_{st} \alpha'^2$. Supergravity can only be trusted if
the length scale of the 3-brane solution is much larger than the 
string scale $\sqrt{\alpha'}$,
i.e. for $N g_{st} \gg 1$.
Of course, the incident energy also has
to be small compared to $1/\sqrt{\alpha'}$. 
Thus, the supergravity calculation should
be valid in the ``double-scaling limit'' \cite{kleb}
\begin{equation}
\label{dsl}
{L^4\over \alpha'^2} = 4\pi  g_{st} N \rightarrow \infty\ ,
\qquad\qquad \omega^2 \alpha' \rightarrow 0\ .
\end{equation}
If the description of the extremal 3-brane background by a stack of
many coincident D3-branes is correct, 
then it {\it must} agree with the supergravity results in this limit.
Since $4\pi g_{st} = 2 g_{\rm YM}^2$, this corresponds to the limit of
{\it infinite} `t Hooft coupling in the  
${\cal N}=4$ $U(N)$ SYM theory.\footnote{Since we also want to send $g_{st}
\rightarrow 0$ in order to suppress the string loop corrections,
we necessarily have to take the large $N$ limit.}

On the other hand, in the gauge theory we have
calculated the absorption
cross-section to leading order in the `t Hooft coupling. So, why
should this result agree exactly with the strong coupling prediction of
supergravity? The answer is that all higher-order corrections
in the coupling cancel out exactly, i.e. the leading order
calculation is protected by a non-renormlaization theorem.
Such a theorem was first enhibited explicitly in the context
of graviton absorption \cite{gkThree}, but since the graviton and
the dilaton belong to the same supermultiplet, supersymmetry should
be sufficient to guarantee that the absorption of the dilaton is similarly 
protected. 
The absorption cross-section for a graviton polarized along
the brane, say $h_{xy}$, is related to the discontinuity accross the
real axis (i.e. the absorptive part) of the two-point function
$\langle T_{xy} (p) T_{xy} (-p) \rangle$
in the SYM theory. In turn,
this is determined by a conformal ``central charge''
which satisfies a non-renormalization theorem: it is completely
independent of the `t Hooft coupling. A simple plausibility argument for 
this fact is that, if there exists a $c$-theorem (and it is indeed
suspected to exist in 4 dimensions \cite{Ans}), 
then the central charge should be constant along critical lines. 

In general, the two-point function of a gauge invariant operator in
the strongly coupled SYM theory may be read off from the
absorption cross-section for the supergravity field which
couples to this operator in the world volume action \cite{gkThree,KTV}.
For a canonically normalized bulk scalar field coupling to the
D3-branes through an interaction
\be
S_{\rm int} = \int d^4 x \phi (x,0) {\cal O} (x)
\ee
($\phi(x,0)$ denotes the value of the field at the transverse
coordinates where the D3-branes are located)
the precise relation is given by
\be
\label{eq:disc}
\sigma = \left. {1 \over 2 i \omega}  {\rm Disc}\; \Pi (p) \right|_{-p^2
=
\omega^2 - i \epsilon}^{-p^2 = \omega^2 + i \epsilon}
\ee
Here $\omega$ is the energy of the particle, and
\be
\Pi(p) = \int d^4 x e^{i p \cdot x} \langle {\cal O}(x) {\cal O} (0)
\rangle
\ee
which depends only on $s=p^2$. To evaluate (\ref{eq:disc}) we extend
$\Pi$ to  complex values of $s$ and compute the discontinuity of $\Pi$
across the real axis at $s=\omega^2$.

Some examples of the field operator correspondence may
be read off from (\ref{sint}). Thus, we learn that the dilaton
absorption cross-section measures the normalized 2-point function
$\langle O_\Phi  (p) O_\Phi (-p) \rangle$ where $O_\Phi$
is the operator that couples to the dilaton: 
\be
O_\Phi = {1\over 4}\tr (F^2 + \ldots )
\ee
(we have not written out the scalar and fermion terms explicitly).
The agreement of this two-point functions
with the weak-coupling calculations performed in
\cite{kleb,gukt} is explained by a non-renormalization
theorem related by supersymmetry to the
non-renormalization of the central charge discussed in
\cite{gkThree}. 
Thus, the proposition that the $g_{\rm YM}^2 N\rightarrow
\infty$ limit of the large $N$ ${\cal N}=4$ SYM theory can be extracted from 
the 3-brane of type IIB supergravity has passed an important consistency check.

It is of further interest to perform similar comparisons for
more complicated two-point functions.
Consider, for instance, absorption of the dilaton in the
$l$-th partial wave which can be extracted from the radial
equation (\ref{partialthree}).
The thickness of the barrier through which the particle has to
tunnel increases with $l$, and we expect the cross-section
to become increasingly suppressed at low energies.
Indeed, a detailed matching calculation \cite{gukt,KTV} gives  
\begin{equation}
\sigma^l_{SUGRA} = \frac{\pi^4}{ 24}  \frac{(l + 3) (l + 1)}{ [(l + 1)!]^4}
\left( \frac{\omega L}{ 2} \right)^{4l} \omega^3 L^8.
\end{equation}
Replacing $L^4$ through (\ref{throatrel}) this can be rewritten as
\begin{equation}
\sigma^l = \frac{N^{l + 2} \kappa^{l + 2} \omega^{4l + 3} (l + 3)}{
3 \cdot 2^{5l + 5} \pi^{5l/2 + 1}l![(l + 1)!]^3}.
\label{eq:semiclassical-absorption}
\end{equation}
What are the operators whose 2-point functions are related to these
cross-sections? For a single D3-brane one may expand the
dilaton coupling in a Taylor series in the transverse coordinates to
obtain the following bosonic term \cite{kleb}:
\be
{1\over 4 l!} F_{\alpha\beta} F^{\alpha\beta} X^{(i_1}\ldots X^{i_l)}
\ ,
\ee
where the parenthesis pick out a transverse traceless polynomial,
which is an irreducible representation of $SO(6)$.
The correct non-abelian generalization of this term is
\cite{KTV}
\be \label{incomplete}
{1\over 4 l!} {\rm STr} 
\left [F_{\alpha\beta} F^{\alpha\beta} X^{(i_1}\ldots X^{i_l)}
\right ]
\ ,
\ee
where STr denotes a symmetrized trace \cite{Tseytlin}: 
in this particular case we have to average
over all positions of the $F$'s modulo cyclic permutations.
A detailed calculation in \cite{KTV} reveals that the 2-point function
of this operator calculated at weak coupling
accounts for ${6\over (l+2)(l+3)} $ 
of the semiclassical absorption cross-section
(\ref{eq:semiclassical-absorption}) in the sense of the relation
(\ref{eq:disc}). Luckily, (\ref{incomplete}) is not the complete
world volume coupling to the dilaton in the $l$-th partial wave.
The ${\cal N}=4$ supersymmetry of the DBI action
guarantees that there are additional terms
quadratic and quartic in the fermion fields \cite{Mark-Wati-4}.\footnote{
In order to
see why there are no more than 4 fermion fields, we note that the
complete operator may be found by acting with 4 supercharges
on the superconformal primary operator ${\rm STr}
[X^{(i_1}\ldots X^{i_{l+2})}]$.} 
When all these terms are
taken into account there is {\it exact} agreement between the weak
and strong coupling calculations of the 2-point functions (for details of 
these rather involved calculations see \cite{KTV}).
This strongly suggests that the complete $l$-th
partial wave operators are protected by supersymmetric
non-renormalization theorems. For recent progress towards
proving such theorems see \cite{Sken}.

\section{Correlation functions and the bulk/boundary correspondence}

Following the circle of ideas reviewed in the previous section, 
Maldacena made
a simple and powerful observation \cite{jthroat}
that the ``universal'' region of
the 3-brane geometry, which should be directly identified with the
${\cal N}=4$ SYM theory, is the $AdS_5\times S^5$ throat, 
i.e. the region $r\ll L$. 
Following this work, methods 
for calculating correlation functions
of various operators in the gauge theory were proposed in \cite{US,EW}.
\cite{US} proposed
to identify the generating functional of connected
correlation functions in the gauge theory with the extremum of the
classical string theory action $I$ subject to the boundary conditions 
that $\phi(\vec x, z) = \phi_0 (\vec x)$ at
$z=z_{cutoff}$ (at $z=\infty$ all fluctuations are required to vanish):\foot{
As usual, in calculating correlation functions in a CFT
it is convenient to carry out the Euclidean continuation. On the string
theory side we then have to use the Euclidean version of $AdS_5$.} 
\be
   W[\phi_0 (\vec x)] = I_{\phi_0 (\vec x)} 
     \ . 
\ee 
 $W$ generates the connected Green's functions of the gauge theory
operator that corresponds to the field $\phi$ in the sense explained
in section 2.2, 
while $I_{\phi_0 (\vec x)} $ 
is the extremum of the classical string action
subject to the boundary conditions (if we are interested in the
correlation functions at infinite `t Hooft coupling, then 
extremizing the
classical string action reduces to solving the equations of
motion in type IIB supergravity).
An essentially identical prescription
was also proposed in \cite{EW} except there the boundary conditions were
imposed at $z=0$.

The physical meaning of the cut-off on the $z$-coordinate
introduced in \cite{US} is that it acts as a UV
regulator in the gauge theory. 
A safe method for performing calculations of correlation
functions, therefore, is
to keep the cut-off on the $z$-coordinate at intermediate stages
and remove it only at the end \cite{US,freed}. This way the calculations are
not manifestly AdS-invariant, however. One may instead look for
a regularization of
the action which is manifestly AdS invariant \cite{KWnew}. Luckily, when
all subtleties are taken into account, these two ways of performing 
calculations do agree \cite{KWnew}.

\subsection{Two-point functions}

Below we present a brief discussion of two-point functions
of scalar operators. The corresponding field in $AdS_{d+1}$ is a scalar
field of mass $m$ with the action
\be
{1\over 2} \int d^{d+1} x \sqrt g
\left [ g^{\mu\nu} \partial_\mu \phi \partial_\nu \phi
+ m^2 \phi^2 \right ]=
{1\over 2} \int d^d x dz  z^{-d+1}\left [(\partial_z \phi)^2+
(\partial_i\phi)^2  + {m^2\over z^2} \phi^2 \right ]
\ ,
\ee
where we have set $L=1$.
In calculating correlation functions of vertex
operators from the AdS/CFT correspondence,
the first problem is to reconstruct an on-shell field in $AdS_{d+1}$
from its boundary behavior. 
The small $z$ behavior of the classical solution is
\be \label{bc}
\phi (z, \vec x)\rightarrow z^{d-\Delta}
[\phi_0 (\vec x) + O(z^2)]
+z^\Delta [A(\vec x) + O(z^2)] \ ,
\ee
where $\Delta$ is one of the roots of
\be \label{relation}
\Delta (\Delta -d) = m^2\ .
\ee
$\phi_0 (\vec x)$ is regarded as a ``source'' function and
$A(\vec x)$
describes a physical fluctuation.

It is possible to regularize the Euclidean
action \cite{KWnew}
to obtain the following value as a functional of the source,
\be
I(\phi_0) =- (\Delta- (d/2)) \pi^{-d/2}
{\Gamma (\Delta)\over \Gamma (\Delta-
(d/2))}
\int d^d \vec x \int d^d \vec x'
{\phi_0 (\vec x) \phi_0 (\vec x')\over |\vec x- \vec x'|^{2\Delta} }
\ .
\ee
Varying twice with respect to $\phi_0$ we find that the two-point
function of the corresponding operator is
\be \label{twopoint}
\langle {\cal O}(\vec x){\cal  O}(\vec x')\rangle=
{(2\Delta - d) \Gamma (\Delta)\over \pi^{d/2} \Gamma (\Delta -
(d/2))} {1\over
|\vec x- \vec x'|^{2\Delta} }
\ .
\ee
Precisely the same normalization of the two-point function follows from a
different regularization where $z_{cutoff}$ is kept at intermediate stages
\cite{US,freed}.

We note that $\Delta$ is the dimension of the operator.
Which of the two roots of (\ref{relation}) should we choose?
Superficially it seems that we should always choose
the bigger root,
\be
\label{dimen}
\Delta_+ = {d\over 2}+\sqrt{ {d^2\over 4} + m^2 }\ ,
\ee
because then the $\phi_0$ term in (\ref{bc}) dominates over the
$A$ term. While for positive $m^2$ $\Delta_+$ is certainly the right
choice (here the other root $\Delta_-$ is negative), it turns out
that for
\be \label{range}
-{d^2\over 4} < m^2 < - {d^2\over 4} + 1
\ee
both roots of (\ref{relation}) may be chosen. Thus, there are {\it two}
possible CFT's corresponding to the same classical AdS action
\cite{KWnew}:
in one of them the corresponding operator has dimension $\Delta_+$
while in the other -- dimension $\Delta_-$. (The fact that there are
two admissible boundary conditions in $AdS_{d+1}$ for a scalar field
with $m^2$ in the range (\ref{range}) has been known since the old work of
Breitenlohner and Freedman \cite{BF}.) 
This conclusion resolves the following
puzzle. $\Delta_+$ is bounded from below by $d/2$ but there is
no corresponding bound in $d$-dimensional CFT (in fact, as we will see,
there are examples of field theories with operators that violate this
bound). However, in the range (\ref{range}) $\Delta_-$ is bounded
from below by $(d-2)/2$, and this is precisely the unitarity bound
on dimensions of scalar operators
in $d$-dimensional field theory! Thus, the ability to have dimension
$\Delta_-$ is crucial for consistency of the AdS/CFT duality.

A question remains, however, as to what is the correct definition
of correlation functions in the theory with dimension $\Delta_-$.
The answer to this question is related to the physical interpretation
of the function $A(\vec x)$ entering the boundary behavior of the
field (\ref{bc}). As suggested in \cite{BKL} this function is related
to the {\it expectation value} of the operator ${\cal O}$.
The precise relation, which holds even after interactions are taken into
account, is \cite{KWnew}
\be
A(\vec x) = {1\over 2\Delta -d} \langle {\cal O} (\vec x) \rangle
\ .
\ee
Thus, from the point of view of the $d$-dimensional
CFT, $(2\Delta - d) A(\vec x)$
is the variable conjugate to $\phi_0(\vec x)$.
In order to interchange $\Delta$ with $d-\Delta$, it is clear from
(\ref{bc}) that we have to interchange $\phi_0$ and $(2\Delta - d) A$. 
This is a canonical transformation which for tree-level correlators
reduces to a Legendre transform. Thus, the generating functional of
correlators in the $\Delta_-$ theory may be obtained by Legendre
transforming the generating functional of correlators in the $\Delta_+$
theory. This gives a simple and explicit prescription for defining
correlation functions of operators with dimension
$\Delta_-$. For the 2-point function, for example, we find 
that the formula (\ref{twopoint})
is correct for both definitions of the theory, i.e. it makes sense
for all dimensions above the untarity bound,
\be
\Delta > {d\over 2} -1
\ .
\ee
Indeed, note that for such dimensions 
the two-point function (\ref{twopoint}) is positive,
but as soon as $\Delta$ crosses the unitarity bound,
(\ref{twopoint})
becomes negative signaling a non-unitarity of the theory.
Thus, appropriate treatment of fields in $AdS_{d+1}$ gives information
on 2-point functions completely consistent with expectations from 
CFT$_d$. The fact that the Legendre transform prescription of
\cite{KWnew} works properly for higher-point correlation functions
was recently demonstrated in \cite{MW}.

\subsection{Chiral operator dimensions in the $T^{1,1}$ model}

Whether string theory on $AdS_5 \times X_5$ contains fields with 
mass-squared
in the range (\ref{range}) depends on $X_5$. The example 
$X_5= T^{1,1}= (SU(2)\times SU(2))/U(1)$ \cite{Romans,KW}, 
turns out to contain such fields, and the
possibility of having dimension $\Delta_-$ is crucial for the
consistency of the AdS/CFT duality. 
In this case
the dual gauge theory is the conformal limit of 
the world volume theory on a stack of $N$ D3-branes placed at
the singularity of the conifold \cite{KW}.
The conifold may be described by 
an equation in four complex variables,
$\sum_{a=1}^4 z_a^2 = 0$.
Since it is symmetric under an overall rescaling of the
$z$-coordinates, this space is a cone. The base of this cone
is precisely the space $T^{1,1}$, and this implies 
that type IIB string theory on $AdS_5\times T^{1,1}$ 
is dual to the infrared limit of the field theory on $N$ D3-branes
placed at the singularity of the conifold. Since Calabi-Yau spaces
preserve 1/4 of the original supersymmetries we find that this 
should be an ${\cal N}=1$
superconformal field theory. 
This field theory was constructed
in \cite{KW}: it is $SU(N)\times SU(N)$ gauge theory
coupled to two chiral superfields, $A_i$, in the $({\bf N}, \overline{\bf N})$
representation
and two chiral superfields, $B_j$, in the $(\overline{\bf N}, {\bf N})$
representation \cite{KW}. The $A$'s transform as a doublet under one
of the global $SU(2)$'s while the $B$'s transform
as a doublet under the other $SU(2)$, and each field has R-charge $1/2$.
One must also add an exactly marginal superpotential
$W=\epsilon^{ij}
\epsilon^{kl}\tr A_iB_kA_jB_l$.

There is a number of convincing checks of the duality between
this field theory and type IIB strings on $AdS_5\times T^{1,1}$.
We will point out one subtle check which is related to our discussion
of operator dimensions. 
The simplest chiral operators were constructed in \cite{KW}:
\be
\tr (A_{i_1} B_{j_1} \ldots A_{i_k} B_{j_k} )
\ .
\ee
The F-term constraints in the gauge theory require that the
$i$ and the $j$ indices are separately symmetrized; therefore, 
the $SU(2)\times SU(2)$ quantum numbers are $(k/2, k/2)$.
The R-charge is $k$ which determines the operator dimensions to be
$\Delta= 3k/2$. This spectrum of quantum numbers and dimensions
of the chiral operators also follows via the AdS/CFT correspondence from the
spectrum of type IIB theory on
$AdS_5\times T^{1,1}$ \cite{KW,Gubser,Ceres}.
Let us emphasize one interesting subtlety:
the dimension of the $k=1$ operators, $3/2$, is below
$d/2=2$. Thus, it must correspond to the smaller root of
(\ref{relation}), $\Delta_-$.
Analysis of the spectrum of type IIB theory on $AdS_5\times T^{1,1}$
\cite{Gubser,Ceres} reveals the presence of a scalar with
$m^2=-15/4$ which is in the range (\ref{range}). For this value
of $m^2$, $\Delta_-= 3/2$ in perfect agreement with the field theory.

\section{AdS/CFT duality in type 0 context}

In the first part of this talk I discussed supergravity approaches to
theories that occur on
stacks of coincident D3-branes of type IIB theory. An interesting
generalization of this work is to consider instead D3-branes of type 0B
string theory \cite{KT}. Since type 0 strings are
obtained by non-chiral
GSO projections
which break all spacetime supersymmetry \cite{DH}, the resulting field 
theories on branes are necessarily non-supersymmetric.
However, type 0 theories have a well-known
and seemingly fatal flaw in that their closed string
spectrum contains a tachyon of $m^2=-2/\alpha'$. 
I will discuss the simplest example of AdS/CFT duality found
in the type 0 context \cite{KTc} 
and argue that it suggests a novel mechanism for
tachyon stabilization.

The type 0 theories have twice as many massless R-R
fields as their type II cousins, hence they
also possess twice as many D-branes \cite{KT}. For example, since 
the type 0B
spectrum has an unrestricted 4-form gauge potential, there are
two types of D3-branes: those that couple electrically to this
gauge potential, and those that couple magnetically.
Very importantly, the weakly coupled spectrum of open strings on
type 0 D-branes does not contain tachyons after the GSO
projection $(-1)^{F_{open}}=1$ 
is implemented \cite{KT,AP,berg}. Thus, gauge theories living on
such D-branes do not have obvious instabilities at
weak coupling. This suggests via
the gauge field/string duality that the bulk tachyon instability
of type 0 theory may be cured as well \cite{KT,KTc}.

Let us review an argument for the tachyon stabilization
in the simplest setting,
which is the stack of $N$ electric and $N$ magnetic D3-branes \cite{KTc}.
For such a stack the net tachyon tadpole cancels so that there
exists a classical solution with $T=0$. In fact, since the stack couples
to the selfdual part of the 5-form field strength, the type 0B
3-brane classical
solution is identical to the type IIB one. Taking the throat
limit suggests that the low-energy field theory on
$N$ electric and $N$ magnetic D3-branes is dual to the $AdS_5\times S^5$
background of type 0B theory and is therefore conformal in the planar
limit \cite{KTc}. This theory is the $U(N)\times U(N)$
gauge theory coupled to 6 adjoint scalars of the first $U(N)$,
6 adjoint scalars of the second $U(N)$, and fermions in the
bifundamental representations -- 4 Weyl fermions in the
$({\bf N}, \overline {\bf N})$ and 4 Weyl fermions in the
$(\overline {\bf N}, {\bf N})$ (the $U(1)$ factors decouple in
the infrared). Such a theory is 
a $Z_2$ projection of the ${\cal N}=4$ $U(2N)$ gauge
theory \cite{KTc}. The $Z_2$ is generated by 
$(-1)^{F_s}$, where $F_s$ is the fermion number,
together with conjugation by
$\pmatrix{ I&  0\cr  0 & -I\cr}$ 
where $I$ is the $N \times N$ identity matrix. 
This is related to the fact that
type 0 string theories may be viewed
as $(-1)^{F_s}$ orbifolds of the corresponding type II theories \cite{DH}.  
In \cite{NS} it was pointed out that,
since $(-1)^{F_s}$ is an element of the center of the $SU(4)$
R-symmetry, this $Z_2$ projection of the ${\cal N}=4$ theory
belongs to the class of ``orbifold CFT's'' studied in \cite{KS}. 

The fact that this
non-supersymmetric AdS/CFT duality is a
$Z_2$ quotient of the ${\cal N}=4$ duality \cite{KTc,NS} lends
additional credence to its validity. In particular, general arguments
presented in \cite{ber} guarantee that the field theory is 
conformal in the planar
limit and that planar correlators of untwisted vertex operators
coincide with those of the ${\cal N}=4$ theory.
Since this CFT does not appear to have any instabilities at weak
coupling, it was argued in \cite{KTc} 
that type 0B string theory on $AdS_5\times S^5$ is 
stable for sufficiently small radius. This provides
a simple AdS/CFT argument in favor of tachyon stabilization; it is also an
example of how gauge theory may be used to make predictions about
the string theory dual to it.

However, some important aspects of 
this particular $Z_2$ quotient
are related to the twisted sector of the orbifold and are
not covered by the analysis in \cite{ber}. Since
this theory is dual to the type 0B string, which is tachyonic in flat space,
in the limit where the radius 
of $AdS_5\times S^5$ becomes large in string units
this background is unstable \cite{KTc}. From the CFT point of view it
means that the limit of infinite `t Hooft coupling, which is commonly
discussed in the AdS/CFT context, does not make sense.
Indeed, from (\ref{relation}) we conclude that the dimension of the operator
corresponding to the tachyon field via the AdS/CFT map is
complex for large $\lambda= g_{YM}^2 N$.

On the other hand, it is easy to show that this dimension is real
for sufficiently weak coupling.
Using the 
DBI action for D-branes of type 0 theory it was shown in \cite{KTc}
that this operator is
\begin{equation}
{\cal O}_T= {1\over 4} \Tr F_{\a\b}^2 - {1\over 4} \Tr G_{\a\b}^2 
+{1\over 2} \Tr (D_\alpha X^i)^2 - {1\over 2}\Tr (D_\alpha Y^i)^2 + \ldots
\ ,
\end{equation}
where $F_{\a\b}$ and $X^i$ are the field strength and the 6 adjoint
scalars of $SU(N)_1$ while
$G_{\a\b}$ and $Y^i$ are the corresponding objects of $SU(N)_2$.
This operator is odd under the interchange
of the two $SU(N)$'s which is related to the fact that the tachyon is
a twisted state from the point of view of orbifolding type
IIB string theory by $(-1)^{F_s}$. 
For weak coupling the dimension of this operator
approaches 4, and its anomalous dimension can be calculated
as a power series in $\lambda$.
In \cite{KTc} the leading contribution to the anomalous dimension
of this operators was analyzed (it can be read off from the 
2-loop beta functions), and it was shown that
\begin{equation}
\Delta(\lambda)= 4 + {\lambda^2\over 8 \pi^4}+ O(\lambda^3)
\ .
\end{equation}
Thus, the dimension is real for weak coupling but, according to the 
AdS/CFT correspondence, is complex for strong coupling.
A plausible scenario, therefore, is that, 
as $\lambda = g_{YM}^2 N$ is increased,
a transition happens at a critical value $\lambda_c$ \cite{KTc}.
From the
field theory point of view this transition is probably due to a singularity
of the sum of planar diagrams that determines
$\Delta(\lambda)$ \cite{KTc}. In non-supersymmetric large $N$
field theories such singularities are expected to occur on general
grounds \cite{GTone}.

To summarize the second part of the talk, 
we have suggested that finite radius of
convergence of planar graphs, which is a well-known phenomenon
in large $N$ field theories, has interesting implications in view of
the AdS/CFT correspondence. Namely, if there is an operator whose
dimension is real up to a critical value of the `t Hooft coupling
but becomes complex beyond that value, then the dual AdS phenomenon
is that a tachyon present in the large radius (gravity) limit is
stabilized for sufficiently small radii. It is difficult to study
such backgrounds with string-scale curvature directly, but on the dual
CFT side perturbative calculations are quite tractable.
The perturbative stability of the CFT thus provides an argument for
stabilization of tachyons (this is one of the few results to date
where field theory has been used to make new predictions about string theory),
and finiteness of the radius of convergence of planar graphs suggests
how instability may develop at sufficiently strong coupling.

\section*{Acknowledgements}
I am grateful to S. Gubser, A. Polyakov, W. Taylor,
A. Tseytlin, M. Van Raamsdonk and
E. Witten, my collaborators on parts of the material reviewed in this
talk. This work  was supported in part by the NSF grant PHY-9802484 
and by the James S. McDonnell Foundation Grant No. 91-48.


\end{document}